\documentclass[fleqn,twoside]{article}
\usepackage{espcrc2}

\usepackage{graphicx}
\usepackage[figuresright]{rotating}
\usepackage{amsmath}

\def\gsim{\raise0.3ex\hbox{$>$\kern-0.75em\raise-1.1ex\hbox{$\sim$}}}
\def\lsim{\raise0.3ex\hbox{$<$\kern-0.75em\raise-1.1ex\hbox{$\sim$}}}

\title{Preon models, relativity, quantum mechanics and cosmology (I)}

\author{Luis Gonzalez-Mestres\address{LAPP, Universit\'e de Savoie, CNRS/IN2P3, B.P. 110, 74941 Annecy-le-Vieux Cedex, France}}

\begin{document}

\begin{abstract}
Preons are hypothetic constituents of the standard particles. They were initially assumed to have basically similar properties to those of conventional matter. But this is not necessarily the case: the ultimate constituents of matter may feel a different space-time from that of special relativity and exhibit mechanical properties different from those predicted by standard quantum mechanics. They can also play an important cosmological role (inflation, dark matter, dark energy...). It is even not obvious that energy and momentum would have to be conserved in such a scenario. In this series of papers, we review the subject using the superbradyon model as an example, and suggest new ways to explore possible tests of the preon hypothesis. 
\vspace{1pc}
\end{abstract}

\maketitle

\section{Introduction}

In his December 1979 Nobel lecture \cite{Salamlecture}, Abdus Salam said :

{\it "Einstein knew that nature was not economical of structures: only of principles of fundamental applicability. The question we must ask ourselves is this: have we yet discovered such principles in our quest for elementarity, to justify having fields with such large numbers of components as elementary.}

{\it Recall that quarks carry at least three charges (colour, flavour and a family number). Should one not, by now, entertain the notions of quarks (and possibly of leptons) as being composites of some more basic entities (PRE-QUARKS or PREONS), which each carry but one basic charge ? "} (Here, Salam quotes reference \cite{Salam1})

Actually, building complete models able to describe all the standard particles and interactions was not a trivial task. In \cite{preons}, for instance, we presented a constituent model for quarks and leptons similar to quark models of hadrons. The constituents had spin 1/2 and formed a SU(2) doublet. The lowest-lying composite states were assumed to be s-waves, and Fermi statistics was postulated to select the allowed states. We proposed the use of confined SO(3) quantum numbers similar to the SU(3) of color.

But these models did not really raise the question of the validity of fundamental principles such as special relativity and quantum mechanics. This was done for the first time in our 1995 papers \cite{Gonzalez-Mestres1995}, where a radical change of the space-time structure felt by the new particles (presented as possible ultimate constituents of matter) was considered. The question of the validity of quantum mechanics was also raised, and it was specified that assuming quantum mechanics to hold for the new superluminal particles (called later superbradyons) was just a simplifying assumption.

The complexity of more conventional theories, such as those currently based on strings, seems to justify considering possible constituents of standard particles. It must be reminded that, forty years ago, dual models and strings were a way to describe the dynamics of Regge trajectories for hadrons. In the early 1970's, the dual amplitudes leading to string models of hadrons \cite{Goddard} were interpreted in terms of "fishnet diagrams" \cite{NielsenOlesen} involving quarks and gluons.

A controversy \cite{AGM1,AGM2} on the Pomeron structure in the topological expansion \cite{Top} of hadronic scattering amplitudes can illustrate the basic issue. It was shown \cite{AGM1} that, in the dual model from which the string picture has been initially derived, the intermediate state structure (two heavy resonances) of the single loop graph with two twists exhibited the same kinematical properties as those obtained from equivalent multiparticle intermediate states in multiperipheral phenomenological approaches. This was naturally explained in \cite{AGM1} by assuming these resonances to have a multiperipheral internal structure in terms of gluons, as suggested by the fishnet interpretation. More recent phenomenology \cite{Nachtmann} and lattice studies \cite{Meyer} seem to have brought results reasonably consistent with the basic ideas of the glueball Pomeron pattern developed in \cite{AGM1}.

Nowadays, string theories are being applied to conventional "elementary" particles. Therefore, with the present statuts of standard theory, the quest for elementarity seems to justify considering preon models that would raise the question of the ultimate principles of fundamental applicability. 

It must be noticed that the approximations used by Nielsen and Olesen \cite{NielsenOlesen} to derive dual amplitudes from very high-order Feynmann diagrams were based on a specific kinematical selection : the rest energies of the virtual intermediate particles of the graphs where taken to be systematically much larger than the moduli of their energy-momentum quadrivectors. Thus, no track was kept of the actual space-time structure felt by these constituents of matter. Therefore, fishnet diagrams involving superbradyons as internal lines can in principle lead in the same kind of limit to a string structure for conventional particles compatible with standard special relativity. 

A superbradyonic version of the Nielsen-Olesen pattern may be close to a new form of condensed matter of which conventional particles would be low-energy excitations. Then, string-like models would be among the candidates to describe these excitations. The so-called "extra dimensions" may in this case be just a phenomenological tool to describe local degrees of freedom of the physical vacuum that can manifest themselves through the excitation spectrum. Local gauge transformations and gauge interactions can also be the expression of these degrees of freedom. 

\section{Superbradyons}

To date, Lorentz symmetry violation at the Planck scale is not experimentally excluded \cite{Gonzalez-Mestres2008,Gonzalez-Mestres2009,Gonzalez-Mestres2009bis}. No experimental test seems to have been yet considered of the validity of quantum mechanics, and even of energy and momentum conservation, at scales close to the Planck scale.

The question of the validity of special relativity, quantum mechanics and energy and momentum conservation at the Planck scale does not only concern the properties of standard particles. The real subject is actually the deep structure of matter beyond this scale. At the Planck scale or at some other scale related to a fundamental length $a$, the nature of the constituents of matter can drastically change. 

The superbradyon hypothesis considered since 1995 \cite{Gonzalez-Mestres1995,gonSL1,gonSL2} can provide an example of this new physics. The existence of superbradyons, even as cosmological remnants and candidates to dark matter and dark energy, has not been ruled out experimentally. Superbradyons would be preons with a critical speed in vacuum $c_s$ much larger than the speed of light $c$. $c_s$ can possibly replace $c$ in a new symmetry of the Lorentz type, but other scenarios can also be considered. 

Although standard preon models assumed that preons obey conventional special relativity, there is no compelling reason for such a hypothesis. The situation is similar for energy and momentum conservation. It was also initially assumed that preons carry the same kind of charges and quantum numbers as conventional particles, but this is not really necessary. Standard gauge bosons can be composite objects coupled to dynamically generated quantum numbers and charges.

Contrary to tachyons that obey standard special relativity at the price of imaginary masses and cannot emit "Cherenkov" radiation in vacuum, superbradyons would have positive mass and energy, and spontaneously emit radiation in vacuum in the form of conventional particles. They would also be the constituents of the physical vacuum. 

In our models \cite{Gonzalez-Mestres1995,gonSL2,gonSL3} of Lorentz symmetry violation (LSV), we assume the existence in our Universe of an absolute local rest frame (the vacuum rest frame, VRF). A simplifying hypothesis, although not compelling, is to identify the fundamental scale $a$ with the Planck scale. We also assume energy and momentum conservation to be valid for any practical purpose, at least at the energy and momentum scales considered for phenomenology, including the ultra-high energy cosmic-ray (UHECR) scale. This condition is not required at the Planck scale or beyond. 

In \cite{Gonzalez-Mestres2009bis}, we discussed the possibility that superbradyons do not obey standard quantum mechanics, or that quantum mechanics be violated in an energy-dependent way. A possible energy-dependent violation of energy and momentum conservation similar to that of LSV in the quadratically deformed relativistic kinematics (QDRK, \cite{gonSL2,gonSL3}) was also considered.  

\section{Possible observable consequences of the superbradyon hypothesis}

Whether or not superbradyons would be concentrated around conventional astrophysical sources deserves a closer study that we shall not attempt here in detail. In many possible scenarios, this would be partially the case but not completely, as superbradyons are expected to be weakly coupled to conventional gravitation. However, it is not excluded that superbradyons be strongly attracted by astrophysical objects made of standard matter.

As emphasized in \cite{Gonzalez-Mestres2009bis}, the expected very weak coupling of superbradyons to gravitation may be compensated by the relation $E~\simeq ~m~c_s^2~\gg ~ m~c^2$ between the superbradyon rest energy $E$ and its mass $m$, so that in practice it is possible that superbradyonic matter strongly feels gravitation even if it generates very weak gravitational forces as compared to its total energy. 

The question of whether superbradyons obey standard quantum mechanics, as well as that of the universality of the Planck constant, have been evoked in previous papers \cite{Gonzalez-Mestres1995,Gonzalez-Mestres2009bis,gonSL1,gonSL2,gonSL4}. Superbradyons may have a different value of the Planck constant, $h_s$ , just as $c_s$ is different from $c$, or follow a completely different pattern. If $h_s$ is not zero, it can be smaller or larger than the standard Plack constant $h$ used for conventional matter. It may also happen that superbradyon mechanics follows a completely different pattern.

In all cases, the question of how standard quantum mechanics can be generated from superbradyon dynamics is close to that of the interpretation of quantum mechanics on which a large amount of work has been performed \cite{EPR,QM}. As the question of the interpretation of quantum mechanics remains open, considering a possible superbradyonic origin appears legitimate.

Very small differences between the values of $h$ for different kinds of conventional particles may be a sign of compositeness. To date, the value of $h$ recommended by CODATA {\cite{CODATA} (6.62606896 x 10$^{-34}$ J s) is given with a 5.0 x 10$^{-8}$ standard accuracy. Existing results are all based on low-energy measurements, but the phenomenon can be energy-dependent. 

More generally, energy-dependent violations of conventional quantum mechanics for standard particles can be generated at the Planck scale or at the fundamental length scale $a$. They can then follow patterns similar to those considered for Lorentz symmetry violation in our papers \cite{gonSL2,gonSL3}, leading to possible tests at ultra-high energy (UHE) cosmic-ray energies. Then, data \cite{AugerData1,AugerData2,AugerData3,HiresData1} from AUGER \cite{AUG} and HiRes \cite{Hires} and future experiments can potentially be a source of relevant information similar to that provided on LSV by the same experiments.  

As an example, one can consider energy-dependent uncertainty relations for conventional particles involving different components of space momentum in the VRF. For instance :
\begin{eqnarray}
\Delta (p_x^2) ~ \Delta (p_y^2) ~ \gsim ~\alpha _1 ~ [(p_x^2) ~+ ~ (p_y^2)]^2 \\
\Delta (p_x^2) ~ \Delta (p_z^2) ~ \gsim ~\alpha _1 ~ [(p_x^2) ~+ ~ (p_z^2)]^2 \\
\Delta (p_y^2) ~ \Delta (p_z^2) ~\gsim ~\alpha _1 ~ [(p_y^2) ~ + ~ (p_z^2)]^2 
\end{eqnarray}
Where $x$, $y$ and $z$ design orthogonal space directions and $\alpha _1 $ is a very small constant. Then, a simple calculation shows that the minimal uncertainty for the energy variable $E$ would be given by the equation :
\begin{equation}
\Delta E ~\gsim ~3~ .~ 2^{-3/2} ~ \alpha _1 ^{1/2}~ p
\end{equation}
\noindent 
where $p$ is the momentum. 
For a proton, $\Delta E $ would equal the mass term $\simeq ~ m^2~(2~ p)^{-1}$ in the UHE kinematics at $E ~\approx ~ 10^{20}$ eV if $\alpha _1 ~\approx ~ 10^{-45}$. The value of both $\Delta E $ and the mass term would be $\approx ~ 5 . 10^{-3}$ eV . Then, the same effect would happen for an electron at $E ~\approx ~ 10^{17}$ eV assuming a similar value of $\alpha _1$. 

Such energy uncertainties would in principle lead to the instability of the concerned particles. For comparison, the energy width of the $\Delta $ resonance at $E ~\approx ~ 10^{20}$ eV in standard relativity is $\simeq ~ 10^{-3}$ eV . Smaller values of $\alpha _1$ would still be able to produce cosmological effects. 

Another example of commutation relations among momentum components would be :
\begin{eqnarray}
\Delta p_x ~ \Delta p_y ~ \gsim ~\alpha _2 ~a ~p^2~ |p_z| \\
\Delta p_y ~ \Delta p_z ~ \gsim ~\alpha _2 ~a ~p^2~ |p_x| \\
\Delta p_z ~ \Delta p_x ~ \gsim ~\alpha _2 ~a ~p^2~ |p_y|
\end{eqnarray}
\noindent
naturally generating a significant uncertainty for transverse energy at UHE. In both cases, the original commutation relations may involve parity violation and other similar phenomena. 

A few years ago, Schupp et al. \cite{Schupp} considered possible astrophysical implications of non-commutative field theory, having in mind a picture where space-time would have a continuous commutative description at low energies and long distances, but a non-commutative structure at high energies and short distances. To justify this picture, a possible phase transition was invoked. In the examples considered here, no phase transition is required. Similar to our weakly deformed relativistic kinematic (WDSR, \cite{gonSL1,gonSL2}) approach to LSV, the equations can be the same at low energy and at UHE. Standard field-theoretical patterns are then low-energy limits. Cosmological implications would also be important.

The effects just considered are not necessarily specific to possible underlying superbradyon models, but they reflect the relevance of physics beyond the Planck scale. Also, UHE particle instability is far from being the only observable effet to be expected from this kind of new dynamics. 

Similarly, UHECR experiments naturally yield bounds on patterns and values of parameters for possible violations of energy and momentum conservation. Again, such a phenomenon would most likely be energy-dependent and generated at the Planck scale or at the $a$ scale.  

As suggested in \cite{Gonzalez-Mestres2009bis}, violations of energy and momentum conservation may naturally result, for instance, from a possible composite structure of standard particles leading to local inhomogeneities not accounted for in our description of conventional particles. 

\subsection{The origin of the Universe : a superbradyon era?}

A natural hypothesis, if superbradyons can exist as free particles, would be that the evolution of our Universe at temperatures above $\approx ~10^{28}$ K or even lower, including the Grand Unification scale, may have been dominated by the transition from superbradyons to conventional matter. Conventional particles would be generated through "Cherenkov" radiation and through the formation of bound states. One can then assume that the physical vacuum would simultaneously be formed by superbradyon condensation.

Because of the relation $c_s ~\gg ~c$ , superbradyons would in any case have energy $E$ and momentum $p$ such that $E ~\gg ~p~c$. The transition from superbradyonic to standard matter would then produce a fast increase of pressure in the Universe. Similarly, as we expect superbradyons to be weakly coupled to standard gravity, the same transition would considerably increase the effective gravitationally coupled energy. 

Thus, as suggested in previous papers \cite{gonSL1,gonSL2}, the transition from superbradyonic matter to standard matter in the history of our Universe may provide an alternative to standard inflation and produce the required result below the transition temperatures. 

Possible remnants from the superdradyon era would be candidates to dark matter and dark energy. The physical vacuum would also be expected to be made of superbradyons, and conventional particles would be similar to phonons or solitons in this vacuum. 

Assuming that nowadays a privileged vacuum rest frame (VRF) exists in our Universe, this does not imply that the situation was similar in the original superbradyonic universe. The appearance of the LSV generating this privileged reference frame for vacuum is expected to be related to the superbradyon condensation forming this vacuum, but also to the generation of "ordinary" matter from superbradyons, as the conventional special relativity felt by "ordinary" particles would not be an ultimate fundamental property of matter. 

LSV would then be a natural consequence of the superbradyon hypothesis, just as the symmetry of the Lorentz type for phonons in a solid (with the speed of sound instead of $c$) is only approximate \cite{gonSL1,gonSL2}. A solid has an internal absolute reference frame for phonons and other kinds of excitation, even if special relativity is not violated in the Universe.

"Cherenkov" radiation in our vacuum by superbradyonic remnants would produce cosmic rays at all energies that can possibly be detected by existing experiments, from UHECR (\cite{gonSL2}) to (as suggested in \cite{Gonzalez-Mestres2009bis}) the present observations by PAMELA \cite{PAMELA}, ATIC \cite{ATIC}, Fermi LAT \cite{FERMI}, HESS \cite{HESS} and PPB-BETS \cite{PPB-BETS}. Superbradyons can be at the origin of the observed electron and positron fluxes without necessarily being the main component of the gravitational dark matter.

\section{Possible connection with modified gravity}

Superbradyons naturally lead to LSV patterns of the DRK (deformed relativistic kinematics) type \cite{gonSL1,gonSL2}, as a consequence of the introduction of the fundamental length $a$. This deformation also implies the generation of a series of high derivatives in the effective Einstein-Hilbert action for gravitation.

Our initial suggestion \cite{gonSL2} that models violating Lorentz symmetry may provide alternatives to the standard inflationary scenario seems to have been confirmed by papers considering, in a different framework, the possibility that modified gravity drives inflation \cite{grav,stephens}.

It seems worth noticing that a modified (non-local) theory of gravity following the same basic scheme as that used in \cite{gonSL2} to derive DRK has been found \cite{stephens} to lead to an inflationary dynamics equivalent to chaotic inflation. 

In all cases, a complete description of this phenomenon should also include the basic physics (superbradyons in our case, or any other concept) leading to the nonlocality. The possibility that physics generated beyond Planck scale directly manifests itself below Planck scale seems more general than the superbradyon hypothesis.

In the superbradyon era, the role of deformed gravity will increase as standard matter is generated.  

\section{Conclusion}

Using the superbradyon hypothesis as a guide, we have briefly sketched several theoretical and phenomenological issues that can emerge from preon models. A wide range of possible experimental phenomena and tests can be contemplated.

A detailed discussion of the basic questions raised in this note will be presented in forthcoming papers.


\begin{thebibliography}{99}


\bibitem{Salamlecture} Abdus Salam, Gauge Unification of Fundamental Forces, Nobel lecture, 8 December, 1979.
\bibitem{Salam1} Pati, J. C. and Abdus Salam, ICTP, Trieste, IC/75/106, Palermo Conference, June 1975; Pati, J, C., Abdus Salam and Strathdee, J., Phys. Letters 598, 265 (1975); Harari, H., Phys. Letters 86B, 83 (1979); Schupe, M., ibid. 86B, 87 (1979); Curtwright, T. L. and Freund, P. G.O., Enrico Fermi Inst. preprint EFI 79/25, University of Chicago, April 1979.
\bibitem{preons} L. Gonzalez-Mestres, Phys. Rev. {\bf D23}, 2055 (1981).
\bibitem{Gonzalez-Mestres1995} L. Gonzalez-Mestres, arXiv.org papers astro-ph/9505117, gr-qc/9508054 and astro-ph/9601090.
\bibitem{Goddard} For a review, see for instance : P. Goddard, arXiv:0802.3249 .
\bibitem{NielsenOlesen} H. B. Nielsen and P. Olesen, Phys. Lett. {\bf 32B}, 203 (1970).
\bibitem{AGM1} P. Aurenche and L. Gonzalez-Mestres, Phys. Rev. {\bf D18}, 2995 (1978), Z. Physik {\bf C1}, 307 (1979) and Z. Physik {\bf C2}, 229 (1979).
\bibitem{AGM2} J. W. Dash and C.I. Tan, Z. Physik {\bf C1}, 229 (1979); C.I. Tan, Phys. Rev. {\bf D22}, 1024 (1980). J.W. Dash, Z.Phys. C5, 359 (1980).
\bibitem{Top} G. F. Chew and C. Rosenzweig, Phys. Rev. {\bf D12}, 3907 (1975) ; E. Witten, Nucl. Phys. {\bf B160}, 57 (1979); and references therein.
\bibitem{Nachtmann} See, for instance, O.Nachtmann, arXiv:hep-ph/0312279 , and references therein.
\bibitem{Meyer} See, for instance, H.B. Meyer and M.J. Teper, Phys. Lett. {\bf B605}, 344 (2005).
\bibitem{Gonzalez-Mestres2008} L. Gonzalez-Mestres, arXiv.org paper arXiv:0802.2536, and references therein.
\bibitem{Gonzalez-Mestres2009} L. Gonzalez-Mestres, arXiv.org paper arXiv:0902.0994, and references therein.
\bibitem{Gonzalez-Mestres2009bis} L. Gonzalez-Mestres, arXiv.org paper arXiv:0905.4146, and references therein.
\bibitem{gonSL1} L. Gonzalez-Mestres, arXiv.org paper physics/9704017 .
\bibitem{gonSL2} L. Gonzalez-Mestres, arXiv.org papers astro-ph/9505117, astro-ph/9601090, astro-ph/9606054, astro-ph/9606054, hep-ph/9610474, physics/9702026, physics/9703020, physics/9705032, physics/9709006, physics/9712049, physics/9712056, hep-ph/9905454, physics/0003080, astro-ph/0407603 hep-ph/0510361, hep-ph/0601219 and references quoted in these papers.
\bibitem{gonSL3} L. Gonzalez-Mestres, arXiv.org paper physics/9702026 .
\bibitem{gonSL4} L. Gonzalez-Mestres, arXiv.org: astro-ph/0407603 .
\bibitem{EPR} A. Einstein, B. Podolsky and N. Rosen, Phys. Rev. {\bf 47}, 777 (1935); D. Bohm, Phys. Rev. {\bf 85}, 166 and 180 (1952); C. Kiefer, arXiv:quant-ph/0210152; S. Groblacher et al., Nature {\bf} 446, 871 (2007); V. Singh, arXiv: 0805.1779 .  
\bibitem{QM} See, for instance : G.'t Hooft, arXiv:gr-qc/9903084, arXiv:quant-ph/0212095; M. N. Celerier and L. Nottale, arXiv: quant-ph/0609161; M. Caponigro, arXiv:0811.3877; L. Nottale, arXiv: 0812.0941; M. Blasone et al., arXiv:0901.3907; J.M. Isidro et al., arXiv:0902.0143; L. Marchildon, arXiv:0902.3005;  A. G. Kofman and A. N. Korotkov, arXiv:0903.0671; D. Durr et al., arXiv:0903.2601; C. Bastos et al., arXiv:0904.0400; G.C. Ghirardi, arXiv:0904.0958; G.Domenech et al., arXiv:0904.3476; I. Dotsenko et al., arXiv:0907.0114; F. Caravaglios, arXiv:0907.0125; R. Gambini and J. Pullin, arXiv:0905.4402; C.Wetterich, arXiv:0811.0927, arXiv:0906.4919; and references quoted in these papers.
\bibitem{CODATA} P.J. Morr, B.N. Taylor and D.B. Newell, Rev. Mod. Phys. {\bf 80}, 633 (2008). 
\bibitem{AugerData1} E. Roulet for the Pierre Auger Collaboration, arXiv:0809.2210; C. Hojvat for the Pierre Auger Collaboration, arXiv:0810.3922; S. Vorobiov for the Pierre Auger Collaboration, arXiv:0811.0752; M. Risse for the Pierre Auger Collaboration, arXiv:0901.2525; Silvia Mollerach for the Pierre Auger Collaboration, arXiv:0901.4699; Michael Unger for the Pierre Auger Collaboration, arXiv:0902.3787; Fabian Schmidt, for the Pierre Auger Collaboration, arXiv:0902.4613; The Pierre Auger Collaboration, arXiv:0903.1127 and arXiv:0903.3385; and references given in these papers.
\bibitem{AugerData2} The Pierre Auger Collaboration, arXiv:0906.2189, arXiv:0906.2319, arXiv:0906.2347 and references given in these papers.
\bibitem{AugerData3} R. Ulrich for the Pierre Auger Collaboration, arXiv:0906.4691.
\bibitem{HiresData1} D.R. Bergman, for the HiRes Collaboration, arXiv.org paper arXiv:0807.2814 , and references therein.
\bibitem{AUG} Pierre Auger Observatory, http://auger.org .
\bibitem{Hires} High Resolution Fly's Eye, http://www.cosmic-ray.org/
\bibitem{Schupp} P. Schupp et al., arXiv:hep-ph/0212292 , and references therein. See also J. Wess, arXiv:math-ph/9910013 and J. Madore et al., arXiv:hep-th/0001203 .
\bibitem{PAMELA} O.Adriani et al., arXiv:0810.4995 ; A. Mocchiutti et al., arXiv:0905.2551, and references given in these papers.
\bibitem{ATIC} J. Chang et al., Nature {\bf 456} (2008), 362.
\bibitem{FERMI} The Fermi LAT Collaboration, arXiv:0905.0025, arxiv:0905.0636, arxiv:0907.0373, arxiv:0907.3289 and references quoted in these papers.
\bibitem{HESS} The HESS Collaboration, arXiv:0905.0105, and references therein.
\bibitem{PPB-BETS} S. Torii et al., arXiv:0809.0760, and references therein.
\bibitem{grav} J. Khoury, arXiv:hep-th/0612052; T. Biswas, A. Mazumdar and W. Siegel, arXiv:hep-th/0508194 .
\bibitem{stephens} P. Stephens, arXiv:0908.2787 .

\end{thebibliography}
\end{document}